\documentclass[review]{elsarticle}

\usepackage{hyperref}
\usepackage{caption}
\usepackage{subcaption}
\usepackage{xfrac}

\journal{Astroparticle Physics}

\begin{document}

\begin{frontmatter}

\title{Experimental Investigation of the Nature of the Knee in the Primary
Cosmic Ray Energy Spectrum with the GAMMA experiment}

\author[lebedev]{V.~P.~Pavlyuchenko}

\author[yerphi]{R.~M.~Martirosov\corref{correspondingauthor}}
\cortext[correspondingauthor]{Corresponding author. Tel.: (+374)99 23 01 25}
\ead{romenmartirosov@rambler.ru}

\author[lebedev]{N.~M.~Nikolskaya}
\author[lebedev]{A.~D.~Erlykin}
\author[yerphi]{H.~A.~Babayan}
\author[yerphi]{A.~P.~Garyaka}
\author[yerphi]{H.~S.~Vardanyan}
\author[umich]{L.~W.~Jones}
\author[warsaw]{J.~Kempa}
\author[cern]{B.~Pattison}
\author[bordeaux]{J.~Procureur}

\address[lebedev]{Lebedev Physical Institute, Moscow, Russia}
\address[yerphi]{Yerevan Physics Institute, Brothers Alikhanyan 2, 0036 Yerevan, Armenia}
\address[umich]{University of Michigan, Ann Arbor, USA}
\address[warsaw]{Warsaw University of Technology, Branch in Plock, Poland}
\address[cern]{CERN, Switzerland}
\address[bordeaux]{Centre d'Etudes Nucl\'{e}aires de Bordeaux-Gradignan, France}

\begin{abstract}
We present preliminary results obtained by a novel difference method for the study of the nature of the knee in the energy spectrum of the primary cosmic radiation. We have applied this method to data from the GAMMA experiment in Armenia. The analysis provides evidence for the possible existence of a nearby source of primary cosmic rays in the Southern hemisphere.
\end{abstract}


\end{frontmatter}

\section{Introduction}

Up to an energy of $\sim 3\times10^{15}$ eV, the energy spectrum of the primary cosmic radiation (PCR) is well described by a power law with an index of 2.7. At higher energy the index increases rapidly to 3.1 creating the knee in the energy spectrum. It is now more than 50 years since the knee was discovered \cite{kulikov1958}, but its nature is still the subject of intensive discussions due to its importance for understanding the origin of cosmic rays in general. In spite of many attempts to explain the origin of the knee, none of the proposed explanations is generally accepted. The main reason for this is the difficulty of obtaining a direct experimental evidence for individual PCR sources, caused by the multiple deflections of the charged particle trajectories in the chaotic and regular magnetic fields in the Galaxy. On a large scale, the propagation of the PCR particles is close to a Brownian and can be considered as a diffusive transfer.

At present there are 3 basic astrophysical models describing the behavior of the PCR in this energy range:

\begin{itemize}
\item The diffusion model \cite{ptuskin2007}, in which the knee appears as a result of the increased leakage of particles from the Galaxy with rising energy. Since magnetic fields bend heavy nuclei more than light nuclei, protons leave the Galaxy first followed later by heavier nuclei.

\item The model of limited energy \cite{berezhko2006}, which suggests that the knee reflects the maximum energy to which protons are accelerated in the shells of Galactic supernova remnants.

\item The model of a nearby source \cite{erlykinWolfendale1997}, where the spectrum of particles from this source is superimposed on the smooth Galactic spectrum and creates an excess in the knee region which causes the break of the energy spectrum.
\end{itemize}

\section{The GAMMA experiment}

The present attempt to study the origin of the knee is based on the investigation of the diffusion character of the PCR propagation in the Galaxy, and was carried out using the last 3 years experimental data of the GAMMA experiment. GAMMA is located on Mt. Aragats in Armenia at 3200 m a.s.l. (corresponding to $700 \text{ g}/\text{cm}^2$ atmospheric depth). The geographical coordinates are  $l = 40^{\circ} 28' 12''$ N, $\lambda = 44^{\circ} 10' 56''$ E. The GAMMA array registers extensive air showers (EAS) in the energy range $10^{14}-10^{17}$ eV with the help of the surface and underground scintillation detectors. The detailed description of GAMMA, its technical characteristics and the main data available up to date have been presented in \cite{garyakaAstropp2007,gammaTeamJphys2008,gammaTeamJcontphys2013,martirosovBeijing2011,gallantPune2005}.

EAS with the number of charged particles $N_e > 10^5$, zenith angles of $\theta < 40^{\circ}$ in the laboratory coordinate system and the axes with radius of $R < 60$ m from the center of the GAMMA array are selected for the current analysis. The total number of EAS is 3.382.892 taken over an effective life time of 11544 hours.

During the primary treatment of the experimental data the following characteristics of the registered EAS were calculated:

\begin{itemize}
  \item coordinates X and Y of the shower  axis relative to the center of the GAMMA array;
  \item zenith and azimuthal angles $\theta, \varphi$ in the laboratory coordinate system;
  \item EAS size $N_e$ and number of muons $N_\mu$;
  \item the so-called ``lateral age parameter'' $S$, calculated from the lateral distribution function in the Nishimura-Kamata-Greisen (NKG) approximation;
  \item primary energy $E_0$, calculated by the method described previously \cite{gammaTeamJphys2008} using the EAS parameters $N_e, N_\mu, S, \theta $;
  \item Greenwich arrival time.
\end{itemize}

For each EAS the angular coordinates $\theta, \varphi$ in the laboratory coordinate system were converted to the horizontal astronomical coordinates $\xi, h$ in the following way: $h = 90^{\circ} - \theta$ (instead of zenith angle $\theta$ the height above the horizon has been used) and $\xi = 286^{\circ} - \varphi$, since the ``North'' direction of the GAMMA installation is turned $16^{\circ}$ to the East relative to the real North \cite{gallantPune2005}. The count of $\varphi$ angles was conducted from ``East'' counterclockwise. In the astronomical horizontal coordinate system the count of $\xi$ is clockwise from the South.

The arrival direction for each EAS ($\alpha$ - right ascension, $\delta$ - declination) for the equatorial coordinate system was calculated from the horizontal astronomical system by the standard formulae using as well the geographical coordinate of the installation and the EAS arrival time. In addition, equatorial coordinates for each EAS have been recalculated to the Galactic coordinates ($l$ -– longitude, $b$ –- latitude). The correctness of the recalculation was checked by the astronomical utilities \cite{utilcoor}. The total error of the recalculation from the laboratory system to the Galactic system was no more than 10 angular minutes for the period between 1960 and 2060.

\section{Method for the analysis of the experimental data}
This method is based on two natural assumptions.
\begin{enumerate}
  \item According to many experimental results \cite{guillianPhysRev2007} incoming EAS with primary energies of $10^{14} - 10^{17}$ eV are isotropic to a level better than 1\%. This is due to the presence of numerous sources and to the large-scale diffusion transport of charged particles from the sources to the Earth. It is assumed that under these conditions for a rather big number of registered EAS and not too a large distance between source and the Earth, the contribution of a particular source to the EAS from the source direction will smoothly decrease with the rising angle between the source direction and the direction of the incoming EAS. This is the consequence of the diffusive character of the transport. The maximum contribution is expected from the direction to the source, the minimum contribution -- from the opposite direction. With increasing distance to the source, the angular distribution of the incoming charged particles will become wider and tend towards an isotropic distribution, where the difference of contributions from the source and the opposite direction becomes invisible (the limit to the region of the methods sensitivity). Such an approach can be also applied to the other EAS characteristics that depend on the angle of scattering (for example, the PCR mass composition).
  
  \item It is also assumed that the GAMMA installation operates with the same aperture independently of time of the day and season. This provides the same observational conditions for different directions as the Earth rotates. It is the common requirement for the stable operation of the experimental installation.
\end{enumerate}

The difference method for the test of the knee models at $\sim 3\times 10^{15}$ eV was suggested in \cite{pavlyuchenkoLebedevBull2014}. The difference (more accurately -- the diffusion-difference) method for the analysis of experimental data, assuming the diffusive character of the PCR propagation in the Galaxy, is as follows. The whole celestial sphere in the Galactic coordinates is divided into two (typically unequal) parts: one in the given direction ($l_0, b_0$), the other -- in the opposite direction ($l_0-180^{\circ}, -b_0$). The division is made in a way that the number of events for both samples is the same. The characteristics of the EAS from these two parts of the sky are compared with one another. For both sets of events the experimental distributions of the EAS parameter selected for the analysis (or of the combination of several parameters) are calculated. Since both sets have been reduced to equal conditions, these distributions can be subtracted from one another to study possible differences.

The reduction to equal conditions means taking the same limits of intervals for both distributions and the choice of such an angle $\psi_0$ (or $H_0=\cos \psi_0$) of the spherical cone around the direction ($l_0, b_0$), that the number of events $n$ and $n^{anti}$ for both sets are equal and at $H \ge H_0$ the EAS are coming from a part of the sky centered around the given direction, and at $H < H_0$ -- from the opposite sky part. For EAS with angles ($l,b$)
$$
H = \cos{\psi} = \sin{b_0}\sin{b} + \cos{b_0}\cos{b}\cos{(l-l_0)}.
$$
Taking into account assumptions 1 and 2 it can be said that for $n = n^{anti}$ the observation periods for the two parts of the celestial sphere are equal, and any additional validation of the conditions for the EAS registration efficiency is not required.  In the difference method the common background and the possible methodical errors are subtracted automatically, because they are the same for both sets. The error in the assignment of EAS to the incorrect sample at the boundary region due to the errors in the angle estimations does not matter much. The EAS characteristics are practically similar to each other for close arrival angles and they are subtracted as a common background.

The numerical parameter for the difference of two distributions is $\sfrac{\chi^2}{J}$, where  $\chi^2=\sum\limits_i{\left(\sfrac{\Delta_i}{\sigma_i}\right)^2}$, and $J$ is the number of degrees of freedom. The sum runs over all intervals $i$ of the parameter chosen for this analysis. The difference between the distributions in the interval $i$ is equal to $\Delta_i=m_i - m_i^{anti}$ ($m_i$ and $m_i^{anti}$ being the number of events in the two parts of the sky for the given interval $i$ of the parameter under study). The root-mean-square error of this difference is calculated from the Poisson distribution as
$$
\sigma_i = \sqrt{m_i+m_i^{anti}+1} = \sqrt{n_i+1},
$$
where $n_i$ is total number of events in the interval $i$ over the whole observational sphere. This number does not depend on the given angles ($l_0, b_0$). Such independence of $\sigma_i$ is very important for the comparison of the $\sfrac{\chi^2}{J}$ values to each other when scanning the sky in the search for the direction with the maximum difference between the distributions in the given and opposite directions (the maximum of $\sfrac{\chi^2}{J}$), which then may be interpreted as the direction to a source of PCR.

The equality $n = n^{anti}$ for the installation with a limited scanning sector allows us to investigate the whole sky sphere within the limits of the method's sensitivity, since the values of $\sfrac{\chi^2}{J}$ for the given and opposite directions are equal, because the values of $\sigma_i$ and $|\Delta_i|$ are equal. Only the sign of $\Delta_i$ is changed.

\section{Experimental results}

As an experimental parameter, the EAS age parameter $S$ has been chosen because of its weak dependence on the primary energy and on the EAS incoming angles in the laboratory coordinate system. This is a formal parameter derived by fitting the lateral distribution function in the Nishimura-Kamata-Greisen (NKG) approximation to the detectors response. This parameter is not the ``pure'' age of the longitudinal development of the electromagnetic cascade, which is calculated in the cascade theory, but it is linearly correlated to it. The parameter $S$ is small at the beginning of a shower's development. It is equal to about $1$ at the shower maximum and increases with atmospheric depth. The $S$ distribution has a Gaussian shape around its mean value. The average age of proton primaries is smaller than that for heavier nuclei since the latter start developing higher in the atmosphere.

\begin{figure}
  \centering
  \includegraphics[width=0.6\textwidth]{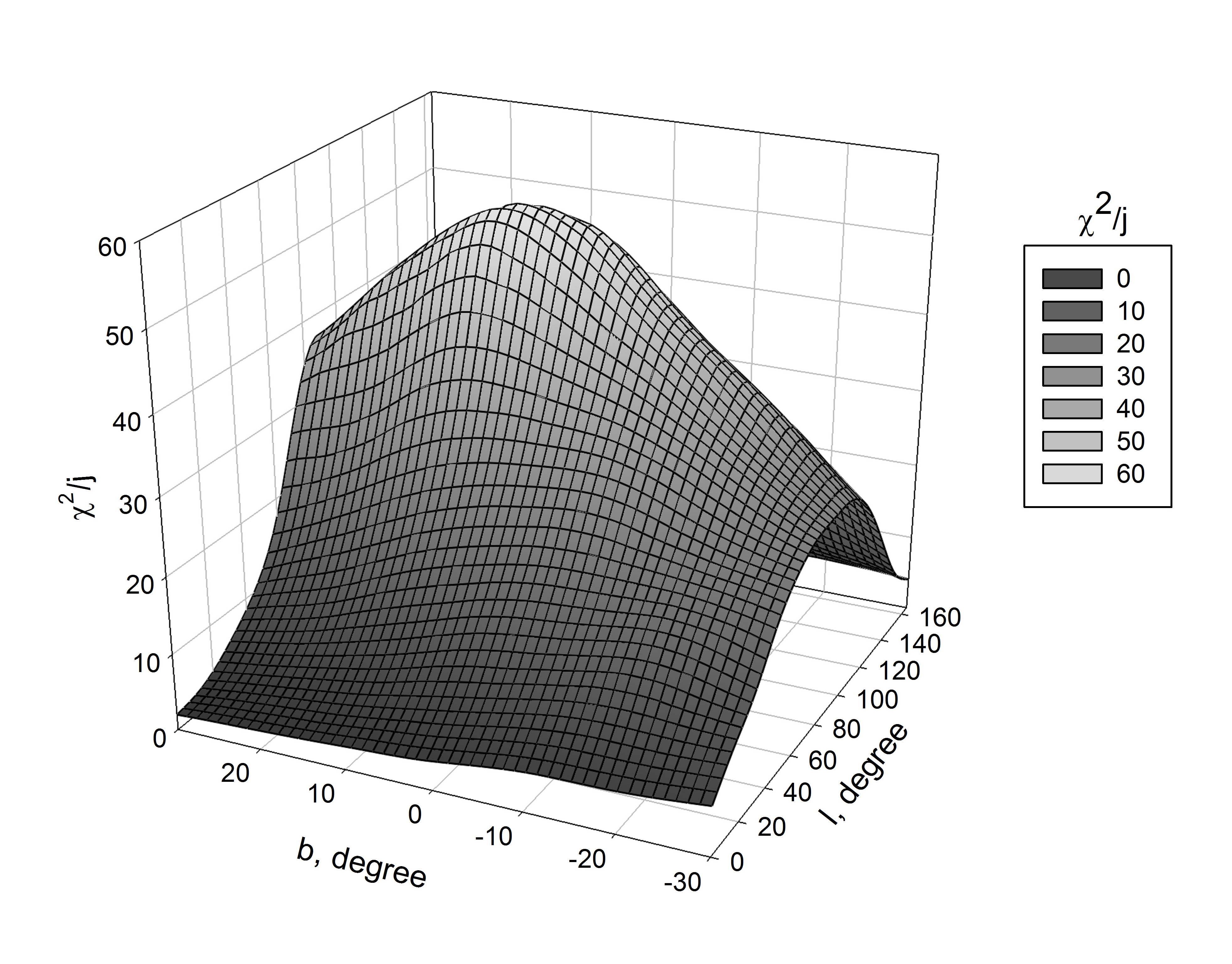}
  \captionof{figure}{$\sfrac{\chi^2}{J}$ distribution for the $S$ parameter in the Galactic coordinate system.}
  \label{Fig1}
\end{figure}

We have scanned the ($l_0, b_0$) plane in order to find the maximum value of $\sfrac{\chi^2}{J}$ (see Fig.~\ref{Fig1}). The range of studied directions is $l_0 = 0^{\circ} \div 180^{\circ}$ and $b_0 = -30^{\circ} \div 30^{\circ}$. The local maximum of the $\sfrac{\chi^2}{J}$ distribution was found in the direction $l_0 = 97^{\circ} \pm 3^{\circ}$, $b_0= 5^{\circ} \pm 3^{\circ}$ (or $l_0 = 277^{\circ} \pm 3^{\circ}$, $b_0= -5^{\circ} \pm 3^{\circ}$ since we get the maximum by comparing opposite directions).

The dependence of the number of EAS in the direction of $l_0 = 97^{\circ}$, $b_0 = 5^{\circ}$ ($H>0.55$) and in the opposite direction ($H<0.55$, $l_0 = 277^{\circ} \pm 3^{\circ}$, $b_0= -5^{\circ} \pm 3^{\circ}$) as a function of the $S$ parameter are presented in Table~\ref{Tbl1}.

\begin{table}[!t]
  \centering
  \begin{tabular}{ | c | c | c | c | c | c | }
    \hline
    $S$ & \begin{tabular}{@{}c@{}}$m_i$ \\ $(H>0.55)$\end{tabular} & \begin{tabular}{@{}c@{}}$m_i^{anti}$ \\ $(H<0.55)$\end{tabular} & $\Delta_i$ & $\sigma_i$ & $\sfrac{\Delta_i}{\sigma_i}$ \\ \hline
0.40	&	1393	&	1634	&	-241	&	55.0	&	-4.4	\\ \hline
0.45	&	3214	&	3617	&	-403	&	82.6	&	-4.9	\\ \hline
0.50	&	6865	&	7650	&	-785	&	120.5	&	-6.5	\\ \hline
0.55	&	14580	&	15870	&	-1290	&	174.5	&	-7.4	\\ \hline
0.60	&	28379	&	30939	&	-2560	&	243.5	&	-10.5	\\ \hline
0.65	&	52522	&	55909	&	-3387	&	329.3	&	-10.3	\\ \hline
0.70	&	88132	&	93369	&	-5237	&	426.0	&	-12.3	\\ \hline
0.75	&	133334	&	138457	&	-5123	&	521.3	&	-9.8	\\ \hline
0.80	&	175895	&	180188	&	-4293	&	596.7	&	-7.2	\\ \hline
0.85	&	205543	&	206558	&	-1015	&	642.0	&	-1.6	\\ \hline
0.90	&	212242	&	210641	&	1601	&	650.3	&	2.5		\\ \hline
0.95	&	198178	&	194725	&	3453	&	626.8	&	5.5		\\ \hline
1.00	&	170464	&	165176	&	5288	&	579.4	&	9.1		\\ \hline
1.05	&	135799	&	131198	&	4601	&	516.7	&	8.9		\\ \hline
1.10	&	103088	&	99050	&	4038	&	449.6	&	9.0		\\ \hline
1.15	&	74108	&	71738	&	2370	&	381.9	&	6.2		\\ \hline
1.20	&	52113	&	50475	&	1638	&	320.3	&	5.1		\\ \hline
1.25	&	35595	&	34252	&	1343	&	264.3	&	5.1		\\ \hline
	&	$<S> = 0.935$	&	$<S> = 0.930$	&	&	&	\\ \hline
	&	$D(S) = 0.156$	&	$D(S) = 0.157$	&	&	&	\\ \hline
  \end{tabular}
\caption{Distributions of the $S$ parameter (column 1) for number of EAS in the direction of $l_0 = 97^{\circ}$, $b_0 = 5^{\circ}$ and in the opposite direction $l_0 = 277^{\circ}$, $b_0 = -5^{\circ}$ (columns 2 and 3). $\Delta_i$ -– difference between the values in columns 2 and 3; $\sigma_i$ -– root-mean-square error of the difference; $\sfrac{\Delta_i}{\sigma_i}$ –- ratio of the difference to its error.}
\label{Tbl1}
\end{table}

\begin{figure}[h]
  \centering
  \includegraphics[width=0.6\textwidth]{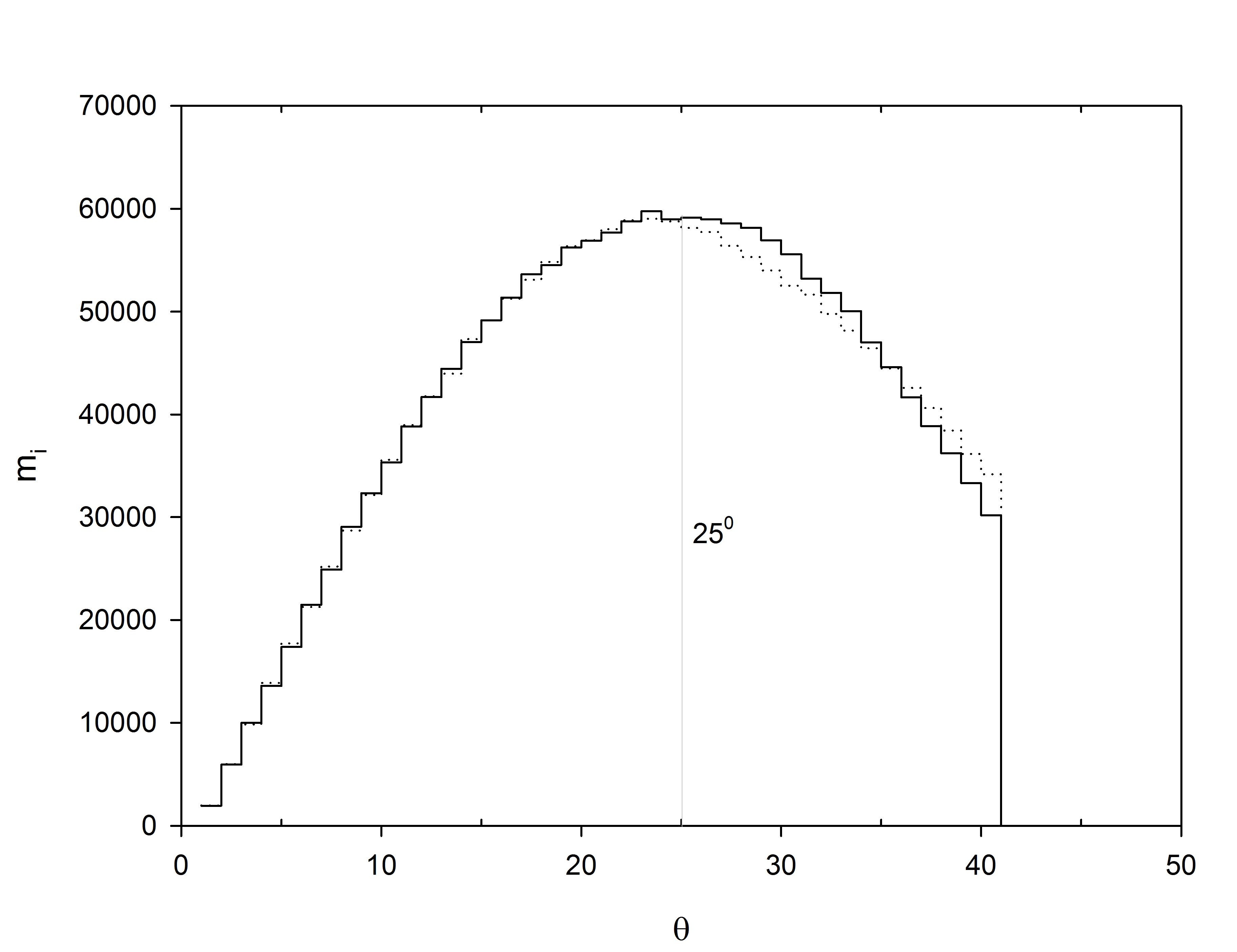}
  \captionof{figure}{Distributions of zenith angles $\theta$ for $H > 0.55$, solid line ($<\theta> = 22.48^{\circ}$) and $H < 0.55$, dotted line ($<\theta> = 22.51^{\circ})$ at $l_0 = 97^{\circ}$, $b_0 = 5^{\circ}$.}
  \label{Fig2}
\end{figure}

In Fig.~\ref{Fig2} we present a comparison of the zenith angle distributions for both regions $H > 0.55$ and $H < 0.55$. This is in order to investigate if the found maximum of $\sfrac{\chi^2}{J}$ might be the result of different zenith angle distributions in both regions of $H$ which possibly could fake this maximum and also the following results. We observe a very good coincidence of the two distributions at $\theta < 25^{\circ}$ but a small difference at $25^{\circ} < \theta < 40^{\circ}$. 

\begin{figure}[h]
    \centering
    \begin{subfigure}[t]{0.5\textwidth}
        \centering
        \includegraphics[width=\linewidth]{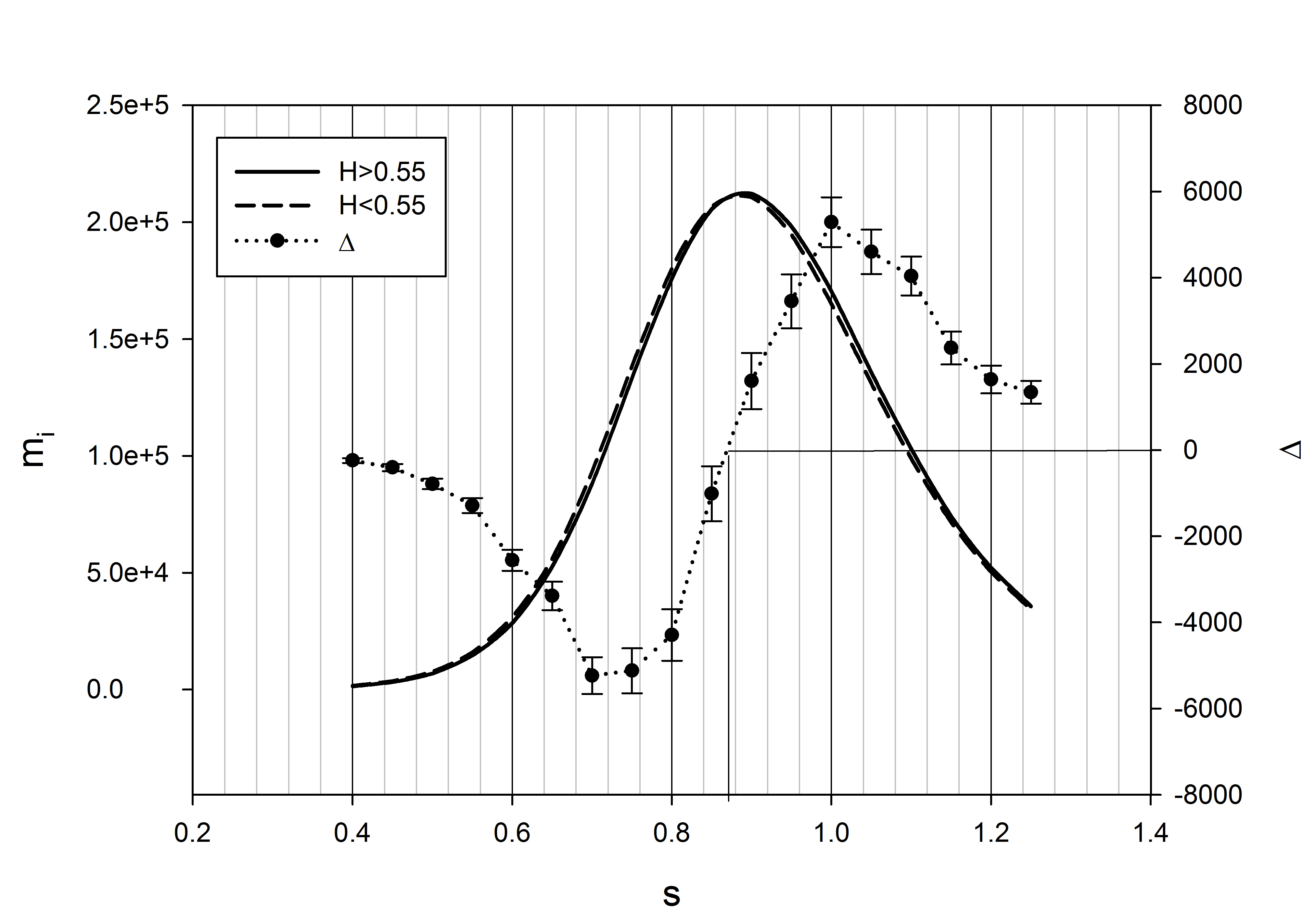}
        \caption{}
        \label{Fig3a}
    \end{subfigure}%
    ~ 
    \begin{subfigure}[t]{0.5\textwidth}
        \centering
        \includegraphics[width=\linewidth]{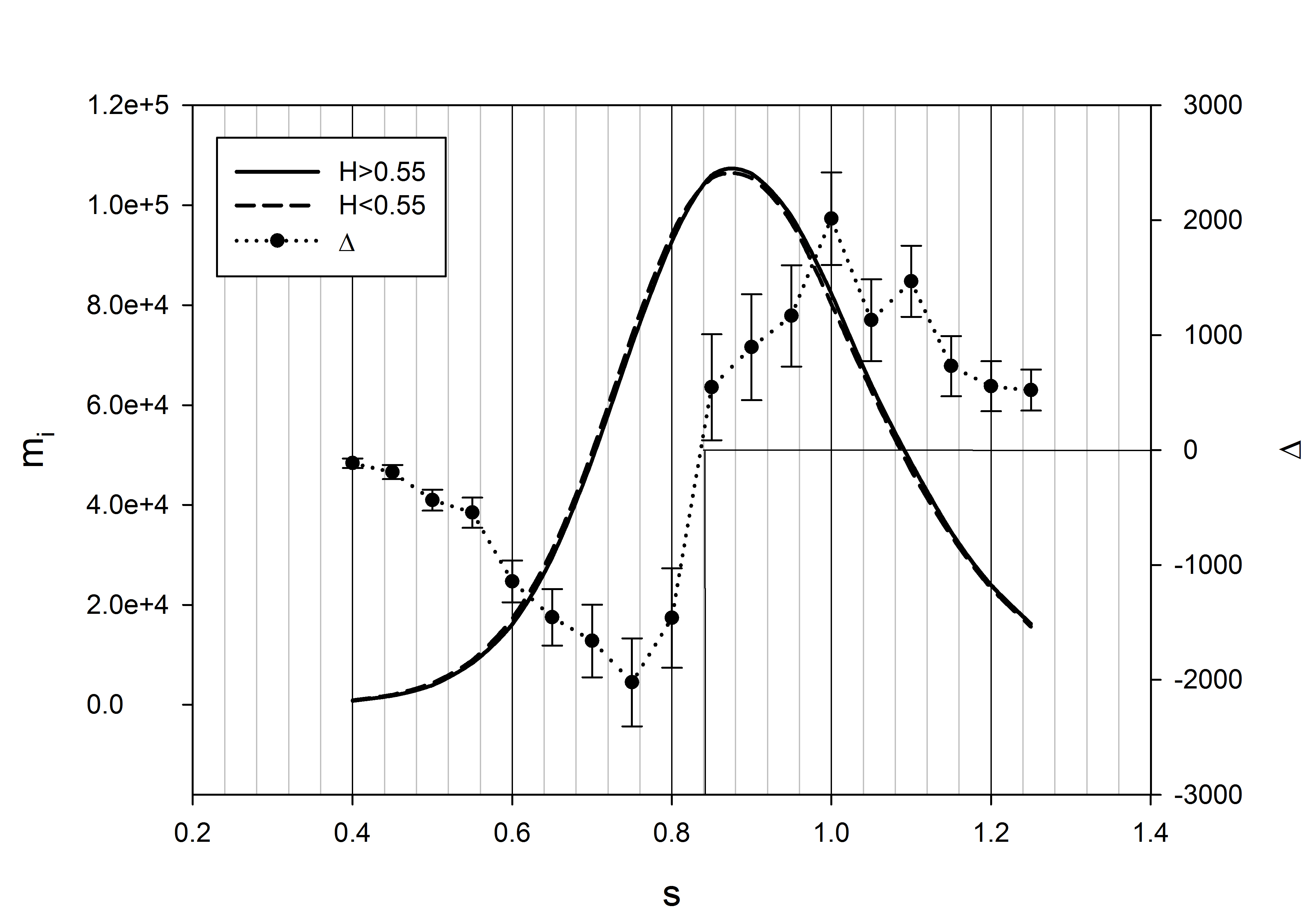}
        \caption{}
        \label{Fig3b}
    \end{subfigure}
    \caption{Number of EAS versus age $S$ for the direction $l_0 = 97^{\circ}$, $b_0 = 5^{\circ}$ and its opposite ($l_0 = 277^{\circ}$, $b_0 = -5^{\circ}$): a) -- for $\theta < 40^{\circ}$, b) -- for $\theta < 25^{\circ}$. The right scales show the difference $\Delta$ between the distributions.}
    \label{Fig3}
\end{figure}

Figure ~\ref{Fig3} shows the $S$ distributions for the numbers of EAS ($m_i$ and $m_i^{anti}$) in the direction $l_0 = 97^{\circ}$, $b_0 = 5^{\circ}$ (solid line), in the opposite direction $l_0 = 277^{\circ}$, $b_0 = -5^{\circ}$ (dashed line) and the difference between these distributions (curve with error bars).

To make sure that the effect seen for the full investigated range of zenith angles (Fig.~\ref{Fig3a}) is not due to the small differences seen for $\theta > 25^{\circ}$ (see Fig.~\ref{Fig2}), we show the distribution also for the restricted range of  $\theta < 25^{\circ}$ (Fig.~\ref{Fig3b}).  The statistics for $\theta < 25^{\circ}$ is two times less than for $\theta < 40^{\circ}$, but the shape of the difference curves as well as the position of the $\sfrac{\chi^2}{J}$ value are the same for both angular ranges within the errors. Therefore the influence of methodical effects connected with different zenith angle distributions in the opposite regions of $H$ is considered negligible. 

We observe a \textit{deficit} of showers with small $S$ for $H>0.55$ which is equivalent to an \textit{excess} of showers with small $S$ for $H<0.55$. This indicates a higher contribution of light nuclei for $H<0.55$ -- i.e. of those nuclei, which have a higher probability to contain any directional information not washed out by diffusion processes. 

\begin{table}[!h]
  \centering
  \begin{tabular}{ | c | c | c | c | c | c | }
    \hline
    $E_0\times10^{-14}$ eV & \begin{tabular}{@{}c@{}}$m_i E_0^{1.7}$ \\ $(H>0.55)$\end{tabular} & \begin{tabular}{@{}c@{}}$m_i^{anti} E_0^{1.7}$ \\ $(H<0.55)$\end{tabular} & $\Delta_i$ & $\sigma_i$ & $\sfrac{\Delta_i}{\sigma_i}$ \\ \hline
		1.00	&	2518		&	2631		&	-113	&	71.7		&	-1.6	\\ \hline
		1.58	&	79831		&	83391		&	-3560	&	597.5		&	-6.0	\\ \hline
		2.51	&	1147168		&	1175618		&	-28450	&	3334.0		&	-8.5	\\ \hline
		3.98	&	4970294		&	5120105		&	-149811	&	10277.9		&	-14.6	\\ \hline
		6.31	&	10005028	&	10268177	&	-263149	&	21548.5		&	-12.2	\\ \hline
		10.00	&	13071512	&	13463556	&	-392044	&	36463.6		&	-10.8	\\ \hline
		15.85	&	13649390	&	13995570	&	-346180	&	55051.0		&	-6.3	\\ \hline
		25.12	&	13156862	&	13367718	&	-210856	&	79762.4		&	-2.6	\\ \hline
		39.81	&	12081955	&	12331834	&	-249879	&	113184.4	&	-2.2	\\ \hline
		63.10	&	10699998	&	10840413	&	-140415	&	157259.0	&	-0.9	\\ \hline
		100.00	&	9177689		&	9380816		&	-203127	&	215904.9	&	-0.9	\\ \hline
		158.49	&	7983836		&	7874125		&	109711	&	295272.1	&	0.4		\\ \hline
		251.19	&	7940497		&	7196120		&	744377	&	426929.7	&	1.7		\\ \hline
		398.11	&	7404421		&	7671228		&	-266807	&	630165.6	&	-0.4	\\ \hline
		630.96	&	7816351		&	6373961		&	1442390	&	906200.9	&	1.6		\\ \hline
$\ge1000.00$	&	13135020	&	13177106	&	-42084	&	1308309.4	&	-0.0	\\ \hline
  \end{tabular}
\caption{Dependence of the parameter $m_i E_0^{1.7}$ on $E_0$ ($m_i$ -– number of events). Column 1 -– $E_0$, columns 2 and 3 -- $m_i E_0^{1.7}$ values in direction $l_0 = 97^{\circ}$, $b_0 = 5^{\circ}$ and opposite direction $l_0 = 277^{\circ}$, $b_0 = -5^{\circ}$. Columns $\Delta_i$, $\sigma_i$ and $\sfrac{\Delta_i}{\sigma_i}$ –- difference between columns 2 and 3, root-mean-square error of the difference and ratio of the difference to its error, respectively.}
\label{Tbl2}
\end{table}

\begin{figure}[h]
  \centering
  \includegraphics[width=0.6\textwidth]{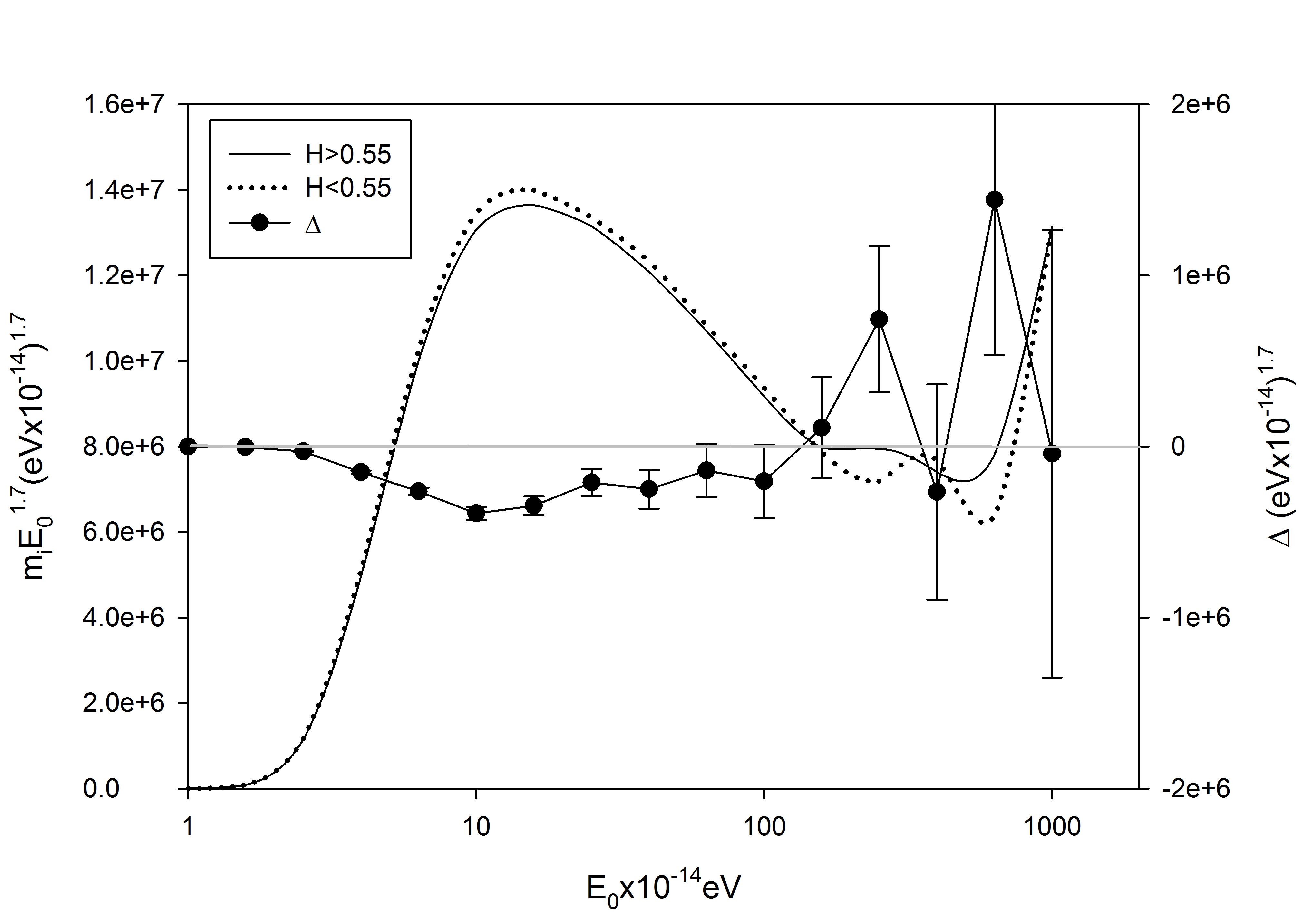}
  \captionof{figure}{Dependence of $m_i E_0^{1.7}$ on $E_0$ for the direction of $l_0 = 97^{\circ}$, $b_0 = 5^{\circ}$ and opposite to it ($l_0 = 277^{\circ}$, $b_0 = -5^{\circ}$). The right scale shows the difference $\Delta$ between the two distributions. }
  \label{Fig4}
\end{figure}

Table~\ref{Tbl2} and Fig.~\ref{Fig4} show the dependence of the parameters $m_i E_0^{1.7}$ and $m_i^{anti} E_0^{1.7}$ on $E_0$. The weighting with $E_0^{1.7}$ has been chosen to highlight details around the knee position and to determine if there is any excess of PCR at $l_0 = 277^{\circ}$, $b_0 = -5^{\circ}$ ($H < 0.55$) in this energy region. Such an excess is expected from Fig.~\ref{Fig3} and the observed small $S$ excess for $H<0.55$, supposing that this region has a stronger contribution of protons than the higher energies where iron PCR are supposed to dominate.  Indeed we observe an excess of PCR in the knee region and slightly above.

\section{Discussion}

The $S$ distributions for the opposite directions are very similar, but the parameter $\sfrac{\chi^2}{J}$ indicates a marked difference. The maximum value of $\sfrac{\chi^2}{J}$ for the points $l_0 = 97^{\circ}\pm 3^{\circ}$, $b_0 = 5^{\circ}\pm 3^{\circ}$ ($l_0 = 277^{\circ}$, $b_0 = -5^{\circ}$) is $57.64 \pm 0.34$ with 17 degrees of freedom. This is a very large value. For a random spread $\sfrac{\chi^2}{J}$ should be close to 1, assuming that all terms have the same error $\sigma$ and, correspondingly, the same statistical weight. In our case the $\sfrac{\chi^2}{J}$ values can be distorted; but this distortion should be the same for all directions of ($l_0,b_0$), because $\sigma_i$ in each interval does not depend on the direction.  Moreover, in our case the value $\left(\sfrac{\chi^2}{J}\right)-1$ linearly depends on the total number of events, which is about 3.38 million. For the control the direction where $\sfrac{\chi^2}{J}$ has minimum equal to $1.32 \pm 0.34$ at $l_0 = 15^{\circ} \pm 10^{\circ}$, $b_0 = 60^{\circ} \pm 10^{\circ}$ was found. This value coincides with the random distribution of $\sfrac\Delta\sigma$ within the limits of one standard deviation. This finding is evidence for the lack of systematic distortions and of a large statistical reliability in the directions $l_0 = 97^{\circ}$, $b_0 = 5^{\circ}$ ($l_0 = 277^{\circ}$, $b_0 = -5^{\circ}$). We emphasize that the direction of minimum $\sfrac{\chi^2}{J}$ is perpendicular to the direction of maximum $\sfrac{\chi^2}{J}$ (although determined with a poorer angular precision than the latter).

\begin{figure}
  \centering
  \includegraphics[width=0.6\textwidth]{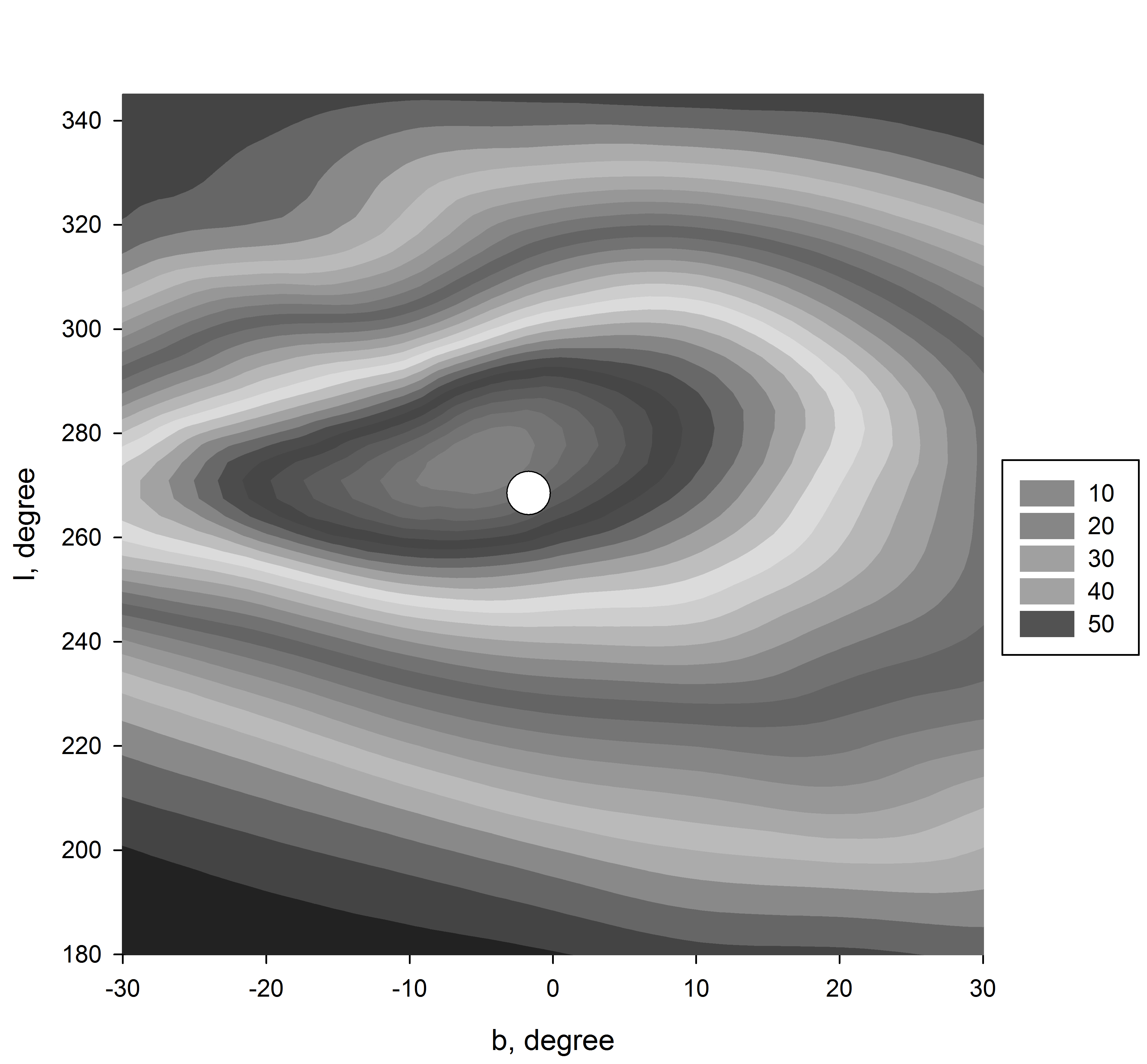}
  \captionof{figure}{$\sfrac{\chi^2}{J}$ distribution for the $S$ parameter in the Galactic coordinate system (contour diagram). The white circle in the center marks the position of the Vela cluster}
  \label{Fig5}
\end{figure}

Table~\ref{Tbl2} and Fig.~\ref{Fig4} demonstrate that there is an excess of EAS in the knee region from the direction of $l_0 = 277^{\circ}$, $b_0 = -5^{\circ}$. Not far from this point there is a cluster in the Vela constellation with two closely appearing  supernova remnants Vela X ($263.9^{\circ}$, $-3.3^{\circ}$) and Vela Jr ($266.2^{\circ}$, $-1.2^{\circ}$) at distances from the Earth of about 0.3 and 0.2 kpc, respectively (Fig.~\ref{Fig5}).

This cluster is a good candidate for being a nearby source of PCR. If we suppose a causal relation between the direction of the observed anisotropy and the Vela cluster we would have to explain the shift in the longitude relative to the supernova remnants and the insufficient axial symmetry. The shift could be connected with possible systematic errors of our analysis or with the existence of a regular magnetic field between the source and the Earth. We emphasize, however, that the main purpose of this paper is to present the diffusion-difference technique as a method to identify tiny anisotropies and to motivate other experimental groups to apply the method to their own data. Given the possible systematics of our analysis we consider it necessary to confirm the effect by other experiments. Anyway we note that Erlykin and Wolfendale recently \cite{erlykinWolfendale2013} draw attention to the fact that Vela could be such a strong local source, if the supernova remnant became ``leaky'' at early times.

Coming back to the rationale of the diffusion-difference method we emphasize that the excess of the ``young'' EAS from this direction may be related to the diffusion of particles on the way from the source to the Earth. ``Younger'' EAS indicate a lighter mass composition of the PCR with predominance of protons. During their diffusion heavy nuclei deviate in the interstellar magnetic field more strongly than light nuclei. That is why the flux of the PCR in the direction of source can be enriched by protons leading to a lighter composition and to the rejuvenation of the EAS coming from the source, compared with the EAS from the opposite side.

It is probable that the registered excess explains the trend to a heavier PCR mass composition at energies above the knee (see \cite{apelAstropp2005,budnevNuclPhys2009}) as well as the decline of the parameter $S$ with rising energy in the region below the knee with its further constancy and, probably, a subsequent increase \cite{cherdyntsevaNuclPhys2003,martirosovBeijing2011}. Such a decline of the $S$ parameter is more rapid than could be expected from the shift of the EAS maximum down with increasing energy.

Based on the obtained preliminary results it is impossible to say, if the excess forms the knee entirely or that the contribution of other sources is possible too, because only an excess was registered, but not an absolute value of the flux from the source.

Only one source is discovered within the radius of the method's sensitivity. Perhaps it is the Vela supernova remnant including sources Vela X and Vela Jr.

Most likely the existence of the diffusion process for the PCR is registered on the way from the nearby source to the Earth. This process initiates a reduction of the mass in the PCR mass composition and, correspondingly, rejuvenation of the EAS in the knee region. The diffusion process in the direction of the Galactic Center -- Anticenter is not observed within the limits of statistical sensitivity of the method.

\section{Conclusion}

We have presented a new method to reveal tiny anisotropies of primary cosmic ray particles, provided they consist of protons and heavier nuclei with different galactic diffusion coefficients. The main feature of the suggested method is a difference study of EAS characteristics but not their intensity in different directions. We have found the age parameter $S$ to be the most suitable and physically motivated parameter. Other parameters may also be useful, and their combination with $S$ may be even more powerful than $S$ alone.

We have used data taken with a comparatively small device, the GAMMA detector in Armenia. We find an anisotropy which is maximal along the direction between the celestial coordinates $l_0 = 277^{\circ}$, $b_0 = -5^{\circ}$ and the opposite sky position. The maximum of the excess at these coordinates turns out to be close to the Vela cluster. The effect has a high statistical significance, but yet we cannot exclude that it is caused by hitherto unconsidered systematic biases. Therefore we suggest that other experiments, with different systematics, repeat the analysis with their own data.

\section{Acknowledgments}

We are grateful to all colleagues at the Moscow Lebedev Institute and the Yerevan Physics Institute who took part in the development and exploitation of the GAMMA array. We are also grateful to the Department of Nuclear Physics and Astrophysics of the Moscow Lebedev Physical Institute, to the Yerevan Physics Institute, to DESY as well as to the State Committee of Science of the Republic of Armenia, to the ``Hayastan'' All-Armenian Fund and Program of Fundamental Research of the Presidium of the Russian Academy of Science ``Fundamental properties of matter and astrophysics'' for financial support.
We also thank Ch.~Spiering and A.~W.~Wolfendale for useful discussions and help in the preparation of this paper.

%\section*{References}
\bibliography{references}

\end{document}